# Can clocks really run backwards?

Charles B. Leffert[*]

5050 Anthony Wayne Drive, Wayne State University, Detroit, MI 48202

**ABSTRACT**

**In an apparently unexplored region of relativistic spacetime, a simple thought experiment demonstrates that conjoined Lorentz transformations predict a proper clock at rest will run backwards and that prediction violates the logical principle of causality. This fault in relativity theory, shown first in a modified version of the standard clock paradox thought experiment, also carries over to cases with finite accelerations of the moving observer. The standard clock paradox, long accepted as showing no paradox, was then re-examined and a logical fault was also found in the concept of spacetime. A relativistic two-dimensional treatment of the Earth's near-circular orbit predicts that our astronomers should measure proper time on distant variable objects in our own Galaxy as impossibly running backward on approach-then-recede trajectories and time running normally forward on recede-then-approach trajectories. Apart from the many successful predictions of relativity theory, these new findings imply that we still have very much new physics to learn about our spatially three-dimensional universe. It is suggested that space is not a freely stretching medium but is something that is substantive and is being produced.**

## PREFACE: What is meant by: "clocks running backward"?

An ordinary clock could be designed so that the hands run counterclockwise with forward time rather than clockwise, but of course, that is not what is meant. What is meant is that which we consider impossible, the hands of an ordinary clock are turning backwards. Even a clock with a pawl and ratchet wheel designed to prevent backward motion, is running backwards. The sand of an hourglass is flowing up instead of down. The chipped paint is returning from the floor and is being bonded again to the clock frame.

     To say a clock is running backwards means that *time* itself is running backwards and past events in the clock's neighborhood are happening again but in reverse order. Such phenomena have never been observed. Actual event sequences *spontaneously* proceed in one direction only – but that constraint is not built into relativity theory.

     The author will show that, for a region of spacetime apparently never before explored except by the author, the Lorentz transformation of relativity theory does predict the impossible, that the mathematical parameter "t", and therefore *time* and *clocks,* run backwards. To compound the unexpected, the very measurements of *stationary clocks* running backwards, are predicted to be made by moving coordinate clocks running *forward* in *moving time* but *backward* in *stationary time,* i.e., moving on a *negative world line*.

     In present physics *time* is just a symmetric mathematical parameter with the symbol t. The author claims there is no one physical *time* of nature to be 'discovered'.

---

[*] C_Leffert@wayne.edu

For this analysis into the beginnings of relativity theory, it is suggested that the reader hold to his subjective asymmetric time, with its ever-present *now*, and enjoy how, not only experiments, but careful thought experiments can reject false theories of nature.

# 1 INTRODUCTION

Relativity theory is now well accepted in the scientific community and has become embedded in our local basic physics and also in the global physics of cosmology and astronomy. At present there are no experiments that contradict, and many that support, its predictions. If relativity theory contains internal logical faults, then they must be in a region of spacetime neither yet explored experimentally nor even carefully analyzed theoretically. A fault in relativity theory could challenge much astronomy.

Nevertheless, even though relativity theory has an excellent record, there is a long-standing conflict between relativity theory and quantum mechanics that has prevented the unification of these two major theories. Some theoreticians believe that it is relativity theory that must somehow be revised (Will 2001).

Relativity theory is basic to several aspects of contemporary astronomy. Astronomers measure the energy and spectrum of radiation from distant sources but many effects attributed to relativity theory can change that radiation on its way from those early emitters to our present detectors. Although motion of the emitter does not change the speed of light, it does change its measured frequency. Gravity and motion of the emitter can change its spectrum. Expansion of space itself decreases the energy and therefore frequency of the photons and the relativistic time dilation decreases their arrival rate. On a different level of analysis, the astronomer must account for the rate of expansion of space over the lookback time and often turns to a relativistic big bang cosmological model for its predicted rate.

Einstein's basic concepts of clock time and non-substantive space are being challenged by quantum mechanics and especially by its virtual particles and by its predicted vacuum energy density of a factor some $10^{120}$ times the astronomer's value.

After the Introduction of Section 1, attention will be directed all the way back to Einstein's 1905 paper (Einstein et al. 1923a) to review his basic concepts of space and time and his solution of how, in principle, to measure coordinate values of distance and time in his vision of space (later spacetime).

In that first paper on the special theory of relativity (SR), Einstein introduced a thought experiment for two reference frames with one having an accelerated motion that was the forerunner of many later thought experiments called the "clock" or "twin" paradox. Section 3 is the heart of the paper and uses various versions of the clock paradox thought experiment to demonstrate and analyze the fault in the theory. The final Section 4 states conclusions and also suggests a path to correct our fundamental concepts. References are given to published works of the author on that path to unification.

**1.1 Space: Free Expansion or Substantive?**

A substantive space is a self-subsistent entity that exists. As conceived here it can be produced as unit cells of constant volume. However, the prevailing concept is a space that can just expand without limit, and is called here a non-substantive space.

One can drive a stake into the ground and consider it a fixed reference point for measurements on the surface of the Earth. With present concepts, there are now no such markers one can, even in principle, plant in space. So how are we to conceive of measurements of time and distance in such a space? Einstein's solution to this problem was given in his 1905 paper (Einstein et al. 1923a). Of course he did not know then that our universe was expanding nor did he know this even many years after he derived his general theory of relativity (GR). His early mindset on space is indicated in this quote from his book: *Relativity* (Einstein 1916):

> Moreover, what is meant here by motion "in space"? From the considerations of the previous section the answer is self-evident. In the first place we entirely shun the vague word "space," of which, we honestly acknowledge, we cannot form the slightest conception, and we replace it by "motion relative to a practically rigid body of reference."

He set up a system of ideal rods and ideal clocks as a representation of space and time on which to base his theory. Standard rigid rods transported from the stationary frame "A" to moving frame "B" served his purpose even though it is predicted that A measures them contracted in B and B measures the A-rods contracted.

Our space is three-dimensional (3-D), so a rigid coordinate system stationary to an observer, could be conceived as consisting of three orthogonal axes (X, Y, Z) with rigid rods lying end to end along each of the three axes with a good *proper clock* at the origin and good *coordinate clocks* placed as needed in this *reference frame*. For the one-dimensional thought experiments considered here only clocks on the X-axis are needed. Some authors find it useful to imagine all 3-D space of interest filled with a framework of such parallel rods and clocks, sometimes called a 3-D *latticework* of rods and clocks (Taylor & Wheeler 1963). For the thought experiments considered here a minimum of six clocks are needed for three reference frames.

The laws of motion of accelerated objects and forces between free particles are legitimate from such a framework but in general the acceleration of a framework itself could be detected so Einstein limited the development of his special theory to deriving relations between rigid frameworks in relative uniform motion of translation, called *inertial frames* where there are no forces acting on the fames. Inertial frames are said to be in free fall, but so are the astronauts in orbit around the Earth. If an instrument to measure a tide-producing gravity reads negative, is the frame inertial? An orbit around a planet is evident from a view out the window but not from a similar orbit around a sleeping black hole in a void with a heading locked on a distant star.

It is the abstract concept of *time* that Einstein decided to mold according to his vision of uniform motion in a gravity-free, non-expanding *space*. The proper *time* of an event "*e*" in space he defined as the reading of a clock at the same (nearby) position as the event *e*. For the development of the theory we assume there is a coordinate clock present at an event *e* for each inertial frame of interest and those coordinate clocks at event *e* will be in the same relative directional motion as their respective origins.

Reflected light ("radar method") can be used to synchronize the clocks at rest in each frame, but from then on, *time* at that position in that frame is assumed to be the reading of that coordinate clock and is assumed to be the same as the proper clock at the origin. For example at event e, the stationary observer may have a coordinate clock "at rest" at e which reads time T and a moving observer may also have a moving coordinate clock momentarily near e that reads a time different than T. To question the difference in clock readings is not to question the clocks but the theory unless, and this is important, the theory forced a moving coordinate clock to a reading by a previous impossible world line.

We represent the successive events of the existence of an entity by a *world line* in our space-time drawings even though there are no traces of that existence in real space. Evidence of events can be recorded on the measuring rods and clocks that we *position* to intercept world lines. In Fig. 1 we could make the B.1 frame the "at-rest" frame and get a different but accurate representation with the A-frame still a straight line at an angle. However, the B.2 frame would also be at a different angle. With B performing a gradual (orbital) turnaround instead of an instant turnaround, and B claiming his frame was inertial, B would plot the A world line as curved. But if A had also moved, could B,s distant measurements account for both contributions to A's curved world line?

Einstein wanted all instruments for measurements of nature in one inertial reference frame to be independent of those in another. He would then derive the equations of transformation that specified the relations of measurements between different inertial frames. Except for $v/C \leq 1$, Einstein set no limit on the relative velocity v between two inertial reference frames, nor did he set a limit on the size of an inertial frame.[*] Such limits will be discussed again in § 4.

In a non-substantive space, each observer uses his own rigid rods for distance. To measure the velocity of the other frame, each observer must use two separated clocks in his framework and make two measurements in passing of a single mark of the other's framework. Also recording two readings of the other's clock at that mark would allow the strange claim by both that the other's clocks are running slower.

The interesting physics is summarized in what is called the *transformation* of coordinate distance and time readings between two reference frames in relative motion. For uniform relative motion, with coordinates (x,t) in the moving frame and coordinates (X,T) in the stationary frame, Newton reasoned that time and space were both independent and absolute and accepted the Galilean transformation for mechanics where lapsed time would be the same ($\Delta T=\Delta t$) and distance could be expressed linearly in terms of velocity and lapsed time $\Delta X=\Delta x+v\Delta t$). Distance is a function of time but time is not a function of distance. Newton did not know that moving clocks slow down and, of course, he did not know the future Maxwell equations for electromagnetism.

**1.2 The Lorentz Transformation**

Einstein did not accept two different transformations between inertial frames for mechanics and electromagnetism so he set about to find one transformation for both. In his 1905 paper with two assumptions: a) "*the principle of relativity*"; that physical laws should be the same in all inertial reference frames and b) "*the constancy of the velocity of light*" C in such frames, he derived the Lorentz transformation (Table 1) where distance is still a function of time but now time is also a function of distance. Maxwell's equations are covariant to it. With this new special theory of relativity, motion in mechanics must also follow the same transformation, so moving clocks run slower, time is a function of position and simultaneity becomes a relative concept depending on the reference frame. Another question arises as to whether this special theory of relativity excludes all accelerations of a reference frame?

---

[*] Locally, *inertial frames* are a useful concept. But, even in the absence of gravity, on cosmic lengths in an expanding universe, the concept runs into trouble, as does the constancy of the velocity of light C.

In the 1905 paper on SR, Einstein describes a "peculiar consequence" of multiple-conjoined (connected sequentially in time, with or without acceleration) inertial systems using thought experiments. Starting with synchronous clocks at positions *a* and *b* in a stationary system, if one of two clocks at position *a* is moved at uniform velocity v along line *ab* to *b*, it will lag the stationary lapse time t by $1/2 tv^2/C^2$ (negligible acceleration). This conclusion was generalized to any polygonal line connecting position *a* to *b* (some accelerations); then it was generalized to a closed polygon trajectory where *b* is back to *a*, (many accelerations) and then finally, the conclusion was further generalized to a continuous closed curve with constant speed v (continuous acceleration). Although Einstein did not derive the physics for the accelerated circular reference frame he did not hesitate to state what that physics must predict i.e., the above time lag $1/2 tv^2/C^2$.

Einstein's circular thought experiment was the precursor for many such thought experiments to follow but it is not one to be performed in a laboratory. The velocity of light is so large, C=1ly/y (i.e., 1 light year per year), that one more realistically reasons in terms of a rocket ship going to the nearest star and returning.

**1.3 Spacetime**

Shortly after Einstein's first paper on relativity theory, H. Minkowski, Einstein's former mathematics professor, in 1906 generalized Einstein's new ideas into a four-dimensional geometry of "spacetime" (Einstein et al. 1923c). Minkowski modeled Einstein's special relativity as a four-dimensional object (manifold) of our three-dimensional space plus *time*. With *time* as a dimension, world lines in this 4-D manifold are assumed complete, that is, spacetime contains no *now*! The Lorentz transformation sets the rules for the inertial frames and the metric. As in Fig. 1 and Fig. 3 for one-dimensional motion, the sheet of paper represents a small section of a 2-D cut through the 4-D manifold.

Selecting a point for the origin and the direction and scale for the perpendicular X and T axes for rest frame A, labels all the points in the manifold. If one now selects a single $(X_e, T_e)$ point and a velocity v (relative to A), then all of the translational constants for the Lorentz transformation equations of Table 1 can be determined and all of the points of the manifold are relabeled (x,t). But the new *distances* and *times* are changed and using the X- and T-axes would distort the 4-D spacetime and any existing world lines.

However, Minkowski showed from geometric considerations (using hyperbola) that one could simply overlay the rest-frame representation with rotated x- and t-axes without displacing existing events and world lines and the transformed values of (x,t) could be read directly from the rotated axes.

The rotated-axes treatment is a very useful tool for analysis of relativistic thought experiments but one must remember the implication of the theory that it is spacetime that is distorted by actual relativistic motion. Some authors describe *spacetime* as "flextime" and "elastic space" (Davies 1995). Einstein finally accepted the concept of *spacetime* and carried it into his general theory where masses also curve spacetime and greatly distort it especially near black holes.

In Fig. 1 the important world lines of both proper clocks are shown and their two intersections that limit the beginning and end of the thought experiment. For clarity of the small figure, only three coordinate clock world lines are shown: two for frame A at 3 light years (ly) and 6 ly and one at –3 ly for frame B. However, tick marks are shown on the X-axis of A where the world lines for the A coordinate clocks from X=-3 ly to X=+7 ly would cross and run parallel to the world line of the A proper clock. There are also two tick marks at x=-1 ly and –2 ly for frame B on the instantaneous x-axis for B coordinate clock world lines that would run parallel to the proper B clock world line.

Note that there are constants at the end of the equations of the Lorentz transformation in Table 1. They are needed for translation of origin for the parallel coordinate clock world lines. Such constants have value zero for proper clocks that pass at a common origin with t=T=0. Neither Einstein nor other relativists mention this detail in their analyses.

For the moving coordinate system, the angle (measured clockwise) from the vertical for the world lines of the moving proper clock B, and its coordinate clocks, is $\theta = \tan^{-1}(v/C)$. For the outbound trip (B.1), $v/C=+0.5$ and $\theta =26.6^o$. The maximum angle for a light ray ($v/C=+1$) is $\theta =45^o$. For the inbound trip (B.2), $v/C=-0.5$ and $\theta =-26.6^o$ and for light $\theta=-45^o$.

The B coordinate clock at x=-3 ly is discussed now for two reasons. First to show consistency and reliability of the Minkowski diagram with the Lorentz transformation and second to show a troublesome feature of spacetime even before discussing the results of the thought experiment.

The x=-3 ly world line demonstrates Einstein's *relativity* of measurements. Note first that on the X-axis (T=0) that the x=-3 ly world line crosses at greater than X=-3 ly showing that A measures B-rods reduced in length. On the other hand if the X=-3 ly tick mark is extended vertically down to cross the instantaneous x-axis (t=0), it crosses at greater than x=-3 ly showing that B also measures A-rods reduced in length.

The instantaneous X-axis moves up in time and remains parallel to its T=0 horizontal position. The instantaneous x-axis, as derived in Table 3, is somewhat more complicated but it too remains parallel to its t=0 angle and moves up in time to the *turnaround* at event *1*. Then with the change in velocity, it rotates from the *1-a* position to the *1-b* position and then, remaining parallel, moves up to the end of the experiment. Since a computer program of the Lorentz transformation is used as a check, one can rely on the Minkowski diagrams to aid in the analysis of the thought experiments.

## 2 CLOCK PARADOX THOUGHT EXPERIMENTS

For our thought experiments in space, we continue with Einstein's approach to measurements in extended (many ly) rigid frameworks of rods and clocks. We begin with simple one-dimensional experiments in gravity-free space involving two such inertial frameworks A and B where B, say, *rockets* to a nearby star and returns to A.

Visualize two long strings of clocks, i.e., many fixed-distance rockets with all clocks synchronized to their respective proper clock, that are passing close enough that at any event of interest, a clock of each framework photographs its own coordinate values and those of the other passing clock. Each framework has an observer (in rocket ship) stationed at its origin, i.e., *observer A* and *observer B*. We arbitrarily pick the A-framework not to be accelerated, so we call it the "at rest" or "stationary" frame.

The simplest experiment is to have the B framework already in motion relative to A at uniform velocity +v along A's X-axis, pass A when all synchronized clocks in each frame have been set to zero as B passes A on the outbound trip. The coordinates in the A-frame are upper case (X,T) and those in the B frame are lower case (x,t). We are particularly interested in the predicted readings of the proper clocks at the origins in these experiments so we also give their readings the Greek symbols $\tau_A$ and $\tau_B$ where, of course, $\tau_A = T_A$ and $\tau_B = t_B$. Numeric and lowercase letter subscripts will also be used to identify events such as $T_1$, $\tau_{B1}$ and $x_a$. Framework B will recede from A at a uniform velocity +v for a fixed lapse of time $\tau_{B1}$ on B's proper clock, then turn around, approach and pass A again at uniform velocity -|v| to the end of the experiment.

Treatments of the clock paradox thought experiments are not very precisely done in the literature. Most investigators have been satisfied just to show that the Lorentz transformation predicts that the moving observer will also measure the correct greater reading of the A proper clock at the end of the experiment. Some ignore the role played by Einstein's framework of coordinate clocks. Their strategy is to have light rays transfer just proper clock readings in one frame to the other and not transfer the passing coordinate clock reading of the other frame (Davies 1995). The Lorentz transformation (Table 1) requires both. Some even completely ignore what happens during turnaround, which also skips the source of the problems (Taylor & Wheeler 1966; Resnick & Halliday 1992). Except for the author's earlier treatment (Leffert & Donahue 1958), the author knows of no careful analysis of the clock paradox thought experiment.

The problems occur in relativity theory when we conjoin two inertial frames with different velocities and these problems are maximized for a sudden reversal of velocity. The goal here is to examine such details very carefully. Later a more realistic gradual turnaround with a constant moderate acceleration will be adopted but for the first simple thought experiments we will take the limit of a zero time lapse for turnaround. Internal forces would demolish a mechanical clock but, in principle, the present laws of physics do not attribute any direct affect of acceleration on clock rate but only by the following effect of the change in velocity. So how does one observer use his rods and clocks to measure the other's such framework?

The proper clock and each of the coordinate clocks have a window displaying its reading of time and under it is a label of its distance from its origin. Each clock also has a camera triggered to photograph the readings of any clock that passes by and its own label and time reading. This information is both recorded on tape and sent by radio signal to the observer at the origin –not to the other observer. Both observers discuss the results after the experiment.

We are now entering *troubled waters* or rather *troubled spacetime*. Note again in Fig. 1, as derived in Table 3, that as B reverses velocity in the limit of zero lapse time that the *instantaneous x-axis* or "line of simultaneity" for B (Taylor & Wheeler 1966) swings from the *1-a* line to the *1-b* line in zero lapse time of B. Relativists have had no trouble in accepting the predicted jump in the proper A clock ($\tau_{Ab} - \tau_{Aa}$) reading in their mission to show no paradox in the standard clock paradox thought experiment. They simply dismiss it with a statement such as, "B must add a constant to the calculated lapse time of the inertial frames to get the correct age of A" (Taylor & Wheeler 1966; Lowry 1963). However, as the author had published earlier, the problem cannot be dismissed so easily as will be shown below.

There are other troublesome features in Fig. 1. The Lorentz transformation predicts for three years, i.e., during the predicted jump in time of A,s proper clock, observer A will record that every B coordinate clock that passes the A proper clock exhibits the same reading of t=5.2 y. Even more troubling, since physicists relate relativistic time with human body aging, a B-coordinate observer at x = -3ly, would not age at all over those same 3 years

Also consider another feature of the turnaround that is troublesome to the relativists when mentioned and which is never discussed in print. That feature is the necessity of continuous <u>maneuvering</u> of the position of all rods and clocks and the <u>re-synchronizing</u> of all coordinate clocks of the B framework during the turnaround whether it is instantaneous or gradual. Indeed, note for the instantaneous turnaround of Fig 1 that every B coordinate clock must undergo a different <u>finite acceleration</u> similar to the one shown for the x=-3 ly coordinate clock. The troubling feature is that one has to <u>maneuver</u> the very instruments during the experiment on which one depends to give the correct measurements. In a thought experiment one can assume the theoretical predictions can happen providing no physical laws have been violated.

In Fig. 1, the Lorentz transformation predicts that B will measure all events of observer A's life for three years happening all at one time. This unphysical prediction is forced by the kinematics of Einstein's special relativity when two of his inertial systems are conjoined with an instantaneous velocity reversal. A rejoinder that the special theory was not derived to include acceleration has not inhibited the claim of relativists that inclusion is still proof of no paradox. While such a turnaround with mass is unreal, this prediction is a harbinger of difficulties yet to come for more natural modest changes in velocity.

There is another principle in physics that is more sacred than the principle of relativity that must not be violated. That is the principle of causality that the cause must precede the effect in time. In other words, in our material world, time and good clocks can not run backwards. Once actual events happen in the order a,b,c they cannot happen again in the order c,b,a. There is nothing built into the equations of relativity theory to keep a reversal of time from being predicted as will be demonstrated in the following.

## 3 LOGICAL FAULTS IN RELATIVITY THEORY

### 3.1 The Standard Clock Paradox

The Minkowski diagram for the standard clock paradox is shown in Fig. 1 where $|v/C|=0.5$ and the speed of light is one light year per year, C=1 ly/y. As described earlier, observer A is assumed to be *at rest*, so his world line is represented as vertical, with time set to zero at the common origin (t=T=0 at x=X=0). The moving observer B has the outbound leg of his world line at v/C=+0.5, labeled "B.1" and the inbound leg, at v/C=-0.5, labeled "B.2." The Lorentz factor $\gamma=(1-(v/C)^2)^{-1/2}=1.15$ so with B's retro-thruster set to fire at time $\tau_B=t=6/\gamma=5.2$ y with assumed infinite acceleration, B returns to A at v/C=- 0.5 and passes A again where the A proper clock reads $\tau_A = T=12.0$ y. The Lorentz constants for this *recede-then-approach* A-B system are given in Table 2 together with the *translation of origin* constants for world lines shown, such as for B.2, that do not pass through the common origin.

The process of *measurement*, as outlined by Einstein, was described earlier as one rigid framework of rods and clocks recording the clock readings and labeled positions of the other passing framework of rods and clocks to verify the predictions of the Lorentz transformation of Table 1. For an instantaneous turnaround of B (event *1*), the Lorentz transformation fixes the entire graph of Fig.1 by the following three specifications: $|v/C|=0.5$ and for A; $T_1=6$ y and for B, $t_1=5.2$ y. Thus Fig. 1, i.e., the Lorentz transformation, already gives us the final end-of-experiment (event *2*) answer that $\tau_{A2} = T_2 =12.0$ y and $\tau_{B2} = t_2 = 10.4$ y. With $T_2 =12$ y at $X_2 =0$ ly, Eq. (1.4) immediately gives $t_2=10.4$ y, but how can B <u>measure</u> proper clock A with a greater time lapse $\tau_{A2}=12.0$ y when for every moment of his inertial motion, B was measuring the proper A clock as running slower as predicted by relativity theory? How can B's coordinate clocks perform such magic? The results of measurements over the entire experiment predicted by the Lorentz transformation are shown graphically in Fig. 2.

A rigid framework of rods and clocks, or even a single finite mass particle, cannot reverse velocity without first passing through zero velocity (see Fig. 4). As we saw in the earlier discussion, and evident in Fig. 1, conjoining inertial frame B.2 to inertial frame B.1, also conjoined all B.2 coordinate clock world lines to those of B.1 and forced all such world lines to display a single value of time, $t= \tau_{B1} = 5.2$ y during turnaround. For B, that feature, in terms of $\tau_B$, is displayed in Fig. 2 by the midpoint vertical time line for B's measurement $\tau_A$ of the A proper clock reading from event *a* to event *b*.

Also when v passes through zero, B must measure the distance to A as $x_A=-3$ ly, that is, more distant than $x_A=-2.6$ ly during the inertial motion of $|v/C|=0.5$ as in Fig. 1. For B, that feature is displayed in Fig. 2 as the midpoint (up-and-down) vertical line of B's measured distance $x_A$ to the A proper clock [For graphical clarity, $-x_A$ is plotted.]. From experience we attach *forces of acceleration* with changes of velocity. There is no term for *acceleration* in the Lorentz transformation. Thus these predicted strange effects have nothing to do with the usual notion of acceleration of a clock, but with a *relativistic change in velocity* between two conjoined inertial reference frames.

At the end of the experiment in Fig 2, when B's proper clock reads $\tau_B=10.4$ y, Eq. (1.2) predicts $\tau_A=12.0$ y. Thus for the standard clock paradox thought experiment, the theory is mathematically consistent in the **<u>end-point</u>** predictions and the paradox is said to be avoided. A mirror-image experiment has B moving on the negative X-axis.

So what is going on here in the change in velocity of this thought experiment and is it acceptable? Lets dig a little deeper. A red flag should have gone up in previous clock paradox analyses (Taylor & Wheeler 1966) when, after derivation of the rotation of the line of simultaneity, one could instinctively guess, just from a graph like Fig. 1, the unphysical predictions that would occur if observer B also measures a friend of A at rest with respect to him such as F at X=6 ly in Fig. 1.

## 3.2 The Modified Clock Paradox or "Causality Paradox"

The world lines for the modified clock paradox thought experiment are shown in the Minkowski diagram of Fig. 3 where C=1 ly/y. The world line of A's friend F is shown at X=6 ly and F measures B at negative values of X. Moving observer B now also measures F at positive values of x. The constants for this *approach-then-recede* B-F system are given in Table 2.

    First note that when B's line of simultaneity gets to line *1-a* at the beginning of turnaround, its extension to F has already passed events *i* and *j* and is now at event *k* on world line F. Then as B's line of simultaneity rotates from line *1-a* to *1-b*, its extension now moves back down world line F and for the second time passes event *j* and stops at event *i*. Then as B's line of simultaneity continues upward, the extension passes event *j* for the third time and event *k* again. Therefore the entire spacetime wedge *i-1-k* on to infinite X is triple valued as the author published some 43 years ago (Leffert & Donahue 1958). The crucial fact however, is that during turnaround B measures events *i, j* and *k* in reverse order and thus he measures the proper F-time and F-clock as running backwards. Such a prediction of any theory is unacceptable because it violates causality and the sequence of events observed by F.

    For the standard clock paradox thought experiment during turnaround when $v_B=0$, moving observer B had to measure the A-proper clock at $x_A=-3$ ly. Here in Fig. 3 during turnaround, B must measure F at $x_F=+3$ ly. To show how Einstein's framework of rods and clocks force that to happen, the world line for B's coordinate clock at x=+3 ly has been added and it does indeed touch the F-world line at $v_B=0$ but only by unacceptably traveling backwards in time – a "negative world line".

    As shown in Fig. 4, the Lorentz transformation generates world lines for B's coordinate clocks at x=-3.0 ly and x=+3.0 ly which are the end points of eleven lines of simultaneity for intermediate velocities from v/C=+0.5 to v/C=-0.5 during turnaround.

    The graphical record of predicted measurements for both observers B and F over the entire modified experiment is shown in Fig. 5. Clearly shown are the triple values of $\tau_B$ for event *j* at $\tau_F=6$ y and the unacceptable reversal in order of events *i, j, k* as $\tau_F$ is predicted to run backwards ($d\tau_F/dt<0$).

    In the standard clock paradox thought experiment of Fig. 2, to satisfy the mathematics and principle of relativity, observer B was forced to measure at turnaround an instantaneous jump <u>forward</u> of $\Delta\tau_A=+3.0$ y in the reading of the A-proper clock that takes $\Delta T = +3$ y to happen. Here in the modified clock paradox thought experiment at turnaround, for the same reasons, observer B is forced to measure an instantaneous jump <u>backward</u> of $\Delta\tau_F = -3.0$ y in the reading of the F-proper clock that takes $\Delta T = -3$ y to happen. This is physically impossible and invalidates relativity theory. As with Fig. 1, there is a similar mirror-image experiment where B moves on the negative X-axis and measures F at X=-6 ly.

    The reader can readily verify the time reversal with the Lorentz transformation of Table 1. Use Eq. (2) with $\delta t_o = 0$, x = +2.60 ly and t = 5.196 y. Outbound at v/C = +0.5, $\tau_{Fk} = T = 1.155(5.196 + 0.5 \cdot 2.60) = 7.5$ y and inbound at v/C = -0.5, $\tau_{Fi} = T = 1.155(5.196 -0.5 \cdot 2.60) = 4.5$ y.

    In preparation for the next section, the reader should also note these same values can be obtained from the new powerful Eq. (11) of Table 3 but using only the stationary

variables X, T and v: $\tau_{Fk}$ = 6.0 –(+.5)(-3.0) = 7.5 y and $\tau_{Fi}$ = 6.0 –(-.5)(-3.0) = 4.5 y. The derivative Eq. (13) also accurately predicts when $d\tau/dt$ will be found negative. Setting $\dot{X}$=u=v and $\dot{u}$=acceleration = $\alpha$, that will be when $[v^2 + X \cdot \alpha]/C^2 > 1$. For the standard case, $\alpha$ = -∞ but X>0, so $d\tau_A/dt$ is positive. For the modified case, also $\alpha$= -∞ but X < 0, so $[v^2 + X \cdot \alpha]/C^2 > 1$, and $d\tau_F/dt$ is negative as we found. Both mirror-image cases are also predicted where the signs of both X and $\alpha$ are reversed.

**3.3 Acceleration and Gravity**

Much good physics followed Einstein's development of the special theory of relativity. Already in his second paper of 1905 (Einstein et al. 1923b), Einstein deduced that the energy of emitted radiation equaled $C^2$ times the change in mass of the emitting particle or $E=mC^2$. That pure mathematical reasoning from two simple assumptions could lead to such success apparently led to a higher goal --- a much higher goal. If the laws of physics are independent of the choice of inertial frames, could it not be that one might derive a field equation such that the laws of physics would be the same in any coordinate frame where the metric was a solution of that field equation? But now, in general, one would have accelerations and gravity to worry about.

     Einstein convinced himself that acceleration and gravity were simply two aspects of the same physics and might be related to the curvature of spacetime. Furthermore, if that reasoning was correct then, in principle, one should be able to measure that curvature of spacetime from within any physical reference frame apart from its possible motion. The development of his general theory was a decade-long task and it was made possible by B. Riemann's development of non-Euclidean geometry. Many books have been written on relativity theory and its many successful predictions of local physics.[*]

     But we are concerned now with the simple physics of finite acceleration in these clock paradox thought experiments. Rather early on, in his book: *The Theory of Relativity*, C. Møller developed the full apparatus of the general theory of relativity to treat this case of the clock paradox with finite accelerations including the fictitious gravity and infinitesimal Lorentz transformation (Møller 1955). The author had reviewed this development (Leffert & Donahue 1958) and more recently has shown (Leffert 1999) that Møller's key Eq. (M-151) was equivalent to Eq. (11) of Table 3.

     For the case of one observer considered at rest in spacetime and the other in relative motion, it apparently never occurred to Einstein or anyone else since, that the kinematics of general acceleration are already latent in the coupling of Lorentz transformations as derived in Table 3 for Eqs. (11), (12) and (13). For this case, the huge GR-apparatus of fictitious gravitational fields and infinitesimal Lorentz transformation amount to no more than the simple Eqs. (11) and (12) of Table 3. Further discussion of this development will be given in § 3.6.2.

     Many cases of accelerated motion of observer B have been studied (Leffert 1999). Finite accelerated motion for the standard and modified clock paradox will be presented next and then finally, one case for the 2-D treatment has been added.

---

[*] I reject the global physics and some local physics of relativity theory.

## 3.4 Clock Paradox with Finite Accelerations

With Eq's. (3-11) and (3-12) and a personal computer, it is simple to repeat the thought experiments of Fig. 1 and Fig. 3 but instead of infinite accelerations four periods of hyperbolic motion (constant acceleration |g|) have been added: (1) +|g| to launch outbound, (2) -|g| to stop at the target, (3) -|g| to begin return and (4) +|g| to end at proper clock A.

    For the thought experiment with finite acceleration g, the periods of uniform motion at |v/C|=1/2 are the same (6 y) as for Figs.1 and 3. The four periods of acceleration were set at $\Delta T$=0.5 y which set $|g|$=1.15 ly/y$^2$. The trajectory of moving observer B followed the hyperbolic motion of Eqs. (14) and (15) of Table 3 with the signs and magnitudes of $x_0$, $u_0$ and g set appropriately in the computer program. The total period for B's journey is increased from 12.0 y to 14.0 y according to A.

    Equations (3-11) and (3-12) were used during the periods of acceleration to calculate the reading $\tau$ of the stationary proper clock and the distance $x_A$ to the stationary clock. This treatment is called the "1-D Simple Lorentzian Treatment (1-D SL)" of accelerated motion. Again please note that the treatment of Eqs. (11) and (12) of Table 3 makes no correction to the special theory, but simply generalizes the Lorentz transformation to include acceleration in the absence of gravity and yields for the clock paradox problem exactly the same predictions as Einstein's GR with the added principle of the equivalence of acceleration and (uniform) gravity

    The standard *recede-then-approach* clock paradox thought experiment with finite acceleration g is presented in Fig. 6. It is clear that these curves are consistent with Fig. 2 and when |g| was increased to 16 ly/y$^2$ these curves closely approached those of Fig. 2. The predicted forward jump in proper A time in Fig. 6 is now spread over an appreciable increment of B-time.

    The important *approach-then-recede* curves for the modified clock paradox thought experiment with finite acceleration g are presented in Fig 7. The data were generated with the same computer program used for Fig. 6 and the only change was a shift in origin of $\Delta X$ = -6.54 ly in Table 2. It is clear that these curves are consistent with Fig. 5 and when |g| was increased to 16 ly/y$^2$, the curves closely approached those of Fig. 5. The predicted backward jump in the proper F-time is now spread over an appreciable finite increment of B-time. All of the curves are now single-valued for B.

    One can now make an approximate check of Eq. (3-13) for $d\tau_F/dt$ for the modified clock paradox with $dX_B/dt$=u. In Fig.7 the center point at X=-2.72 ly has velocity v=0, $\gamma$ = 1 and acceleration g = -1.15 ly y$^{-2}$, so from Eq. 3.13, $d\tau_F/dt \approx 1 - (-2.72)(-1.15) = -2.76$. From the computer printout, take the two points in Fig. 7 on either side of the center point: $\Delta\tau_F/\Delta t \approx (6.62-7.38)/(6.29-6.02) = -2.81$ in reasonable agreement.

    It was pointed out earlier that there are two cases where the proper time of the stationary observer is predicted to run backwards, $d\tau_F/dt$<0. In Eq. (13) of Table 3 for the above case, both X and $\dot{u}$ had negative signs. For the other case both X and $\dot{u}$ have positive signs corresponding to observer B making the outbound leg of the trip along the negative X-axis, undergoing du/dT=+g acceleration, and then returning to A while he also measures A's friend F now positioned at X=-6 ly in which case $X_B$=+3 ly at turnaround and moving observer is again predicted to measure the F-proper time to run backwards.

So we see for the "approach-then-recede" case, that the 1-D SL treatment with finite accelerations also predicts the same flaw in the theory as was shown in the SR-theory with an infinite acceleration turnaround. Next we want to see if the general relativistic treatment also predicts the same flaw in the theory that has been missed during all of these many years.

Møller's early treatment of the clock paradox with the full apparatus of the GR-theory (Møller 1955) contained a key Eq. (M151) and, although Møller had not recognized it, the author showed (Leffert 1999) his equation too contained the same seeds of violation of causality for two cases of *approach-then-recede* motion of the moving observer.

**3.5 Other Clock Paradox Studies**

But Møller's GR-treatment was able to handle two-dimensional circular motion that required the further introduction of a vector potential for the fictitious gravity and that was a challenge to the author who had only a simple one-dimensional *Lorentzian (1-D SL) treatment*. So the author decided to try to combine two orthogonal one-dimensional Lorentzian treatments to simulate two-dimensional motion. The fascinating question was what was one to do with two one-dimensional *times* to get a one two-dimensional *time*? The task turned out to be rather simple and it worked beautifully with the simple mean of the two one-dimensional *times*[•] as summarized in Table 4.

With this new *two-dimensional simple Lorentzian treatment (2-D SL)* generalization of the Lorentz transformation, the author was able to obtain the same predicted measurements from the moving frame for circular motion as Møller but without any consideration of a fictitious gravitational field.

Also for Einstein's 1905 circular motion thought experiment, one could now obtain the predicted results from the moving frame as shown in Fig. 8 where, starting from $\theta=0$ on the X-axis, moving observer B makes one complete counter-clockwise revolution at $v/C=0.9$ with $T_0=10$ y. Note that B's measurement of $\tau_E$ equals T, as it should, only when the separation of the clocks is zero at the beginning and end of the experiment and when their radial velocity of separation is zero at maximum separation $r=-2R$.

Einstein predicted for this case that on return to A, the clock would be slow by a factor of {low-v approximation to:} $(\Delta T-\Delta t)/\Delta T = [1-(1-(v/C)^2)^{1/2}] = 0.5641$ here. From the computer data for Fig 8, $\Delta T=30.944-10.0 = 20.944$ y, $\Delta t=13.488-4.359=9.129$ so $(\Delta T-\Delta t)/\Delta T=0.5641$, in agreement.

For this Einstein case there is no obvious problem with B's measurements of the proper E-time since $\tau_E$ is increasing monotonically, i.e., $d\tau_E/dt>0$. To get a prediction of the stationary clock running backward, it is necessary to increase X in Eq. (13) of Table 3. Therefore, the origin was moved out to Einstein's friend F with $DX=DY=-12$ ly. This was indeed enough to produce $d\tau_F/dt<0$ over the "approach-then recede" part of the orbit. The graph was presented elsewhere (Leffert 1999).

We will return to a similar case in § 3.7 with $d\tau_F/dt<0$.

---

[•] The reader should understand that the author has not been trying to fix relativity theory here but rather to dig deeper for an understanding of where the theory and our fundamental concepts have gone astray.

Can we just ignore this $d\tau_F/dt<0$ - *causality fault* in relativity theory or is their some deeper problem with the theory? Let's go back and re-examine the standard clock paradox thought experiment that has always been accepted as "without paradox" by the scientific community.

**3.6 Logical Fault in Spacetime**

**3.6.1 Three Observer Thought Experiment**

Most physicists for the past 85 years did not challenge the solution of the standard clock paradox thought experiment such as given in Fig. 1 and Fig. 2. Without changing any of the predictions of Figs.1 and 2, we will eliminate the acceleration but keep the change in velocity by using separate inertial frames for B.1 and B.2 with no acceleration during the experiment as shown in the sketch of Fig. 9 for the "turnaround".

Observer B.1 makes the outbound trip as B did before but does not fire his retro-rockets at event "1" but keeps on going. Observer B.2, with his framework of rods and clocks, has been approaching A at v/C = -0.5 and passes B.1 at event "1" where the only transfer is knowledge that both of their proper clocks read 5.2 y. Values for the Lorentz transformation are unchanged.

When B.2 passes B.1, both will have coordinate clocks passing at A. Assume that at least these two coordinate clocks have equipment to record their own reading and to photograph the position label and reading of the opposite coordinate clock and the A proper clock in between. At this event all three observers have clocks close in space.

The Lorentz transformation Eq. (1.1) predicts that both coordinate clocks at A will be labeled x=-2.6 ly (X=0, t=5.2 y, $\delta x_0$=0) and both will have photographed the other's reading as t=5.2 y as indicated in Figs. 1 and 2. The logical fault as shown in this *spacetime* representation is that both coordinate clocks <u>at A</u> must photograph the <u>same</u> reading of proper clock A in between and cannot record the predicted impossibility of two different readings of the one A-clock as demanded by the Lorentz transformation of $\tau_{Aa}$ = 4.5 y and $\tau_{Ab}$ =7.5 y. The kinematics of measurement as set by SR has failed and the *spacetime manifold* has just collapsed revealing the second logical fault in the special theory of relativity. Why collapse?

**3.6.2 The Collapse of Spacetime**

To focus attention, first note that in an animation of Fig. 1, where velocity v decays slowly to zero, Eq, (11) of Table 3 predicts that both $\tau_a$ and $\tau_b$ approach, and then become $T_1$ = 6 y as one might expect.

To follow the collapse of the Minkowski geometric spacetime manifold we go back to § 1.2 and the discussion of re-labeling spacetime points when inertial frames are conjoined. We will carefully rebuild the spacetime of Fig.1 in three steps to show how the fault structure is introduced by the Lorentz transformation.

After placing the rest frame and the point $(X_1,T_1)$ on a blank sheet of paper, we first introduce a second rest frame (v=0) at event 1. The new translational constants of Table 1 place its origin at (x=$X_1$, t=$T_0$=0) and Eq. (11) sets $\tau_A$ = $T_1$ to define a horizontal x-axis at t=$T_1$. If now at event 1, we rotate clockwise to an infinitesimally close point to represent the beginning of a second inertial frame B.1 with v/C = +0.5, Eq. (11) rotates $\tau_A$ counterclockwise from $T_1$ to cross the X=0, T-axis at $\tau_a$ = 4.5 y to define the second instantaneous x-axis at $t_2=T_1/\gamma_2$. However in doing so, as v/C increased from 0 to +0.5, it also reset all of the spacetime points over which it rotated to $t=t_2=T_1/\gamma_2$.

Finally, if now at event 1, we rotate counterclockwise to an infinitesimally close third point to represent the beginning of a third inertial frame B.2, with v/C = -0.5, Eq. (11) rotates $\tau_A$ from $T_1$ clockwise to cross the X=0, T-axis at $\tau_b$ = 7.5 y to define the third instantaneous x-axis, and again in doing so, it reset all of the spacetime points over which it rotated, as v/C decreased from 0 to –0.5, to $t=t_3=T_1/\gamma_3$ where $\gamma_3 = \gamma_2$

Thus if we rule out the rotation of the instantaneous x-axis on change of velocity, which caused the impossible prediction above, $\tau_a$ must collapse back to $T_1$ and $\tau_b$ collapse back to $T_1$. The x-axis becomes parallel to the X-axis and the t-axis becomes parallel to the T-axis and the spacetime concept has just collapsed.

But remember, moving clocks do slow down and electromagnetism does obey the Lorentz transformation, so fortunately, we now have a bright red flag warning us to dig deeper and re-examine our basic physical concepts

The finite acceleration version of the standard clock paradox of Fig. 6 looks more reasonable. Does that mean the spacetime collapse occurs only in the limit of infinite acceleration? The answer is "No." For each $\tau_B$ point of Fig. 6 on one side of the center point with a coordinate clock at A, there is another point on the other side of the center, also with a coordinate clock at A, supposedly measuring a different reading of the proper clock A.

It is claimed to date that man has made no measurements of nature that contradict relativity theory. With all of the above problems appearing in the theory of relativity, is that statement really true or could it be possible that scientists are not sufficiently familiar with enough relativity theory to have recognized nature's contradictions?

**3.7 Experimental Contradiction of Relativity Theory**

**3.7.1 The Problem:** The failure of relativity theory in all of the thought experiments so far presented in §3 have been at relativistic velocities (v/C≥0.5) where no comparable experiment could be conducted by a real moving observer B. Even the scientist who accompanied the atomic clock that was flown around the world just confirmed the "A-predictions" but made no "B-measurements" of any objects on Earth that could have checked Einstein's "B-predictions."

Equation. (13) of Table 3 suggests that perhaps the moving observer B may travel at a modest velocity v and still measure a contradiction of relativity theory, $d\tau/dt<0$, if the object measured is at a great distance X.

The Earth and its astronomers are in a near circular orbit, r ~ $1.59 \times 10^{-5}$ ly, around the Sun at a modest velocity of ~30 km s$^{-1}$ (v/C≈$1 \times 10^{-4}$). Astronomers make periodic measurements of a number of variable physical phenomena such as exploding supernova and jets of matter from black holes with distinct evolution over a period of months. Many supernovae Ia have been carefully measured which increase greatly in luminosity over a period of a few days to a maximum and then dim over a period of months.

The plan now is to consider a variable such as a supernova Ia in a stationary galaxy F such that our moving astronomer measures the luminosity of SN-F in the *approach-then-recede* section of the orbit. The first question is whether the distance X to SN-F is reasonable for relativity theory to predict $d\tau_F/dt<0$ and therefore predict the impossible that the supernova is evolving backwards to the state of the original star.

From Eq. (13), we need $[(dv/dt)X]/C^2>1$. The acceleration of the Earth $dv/dt=v^2/r=(1\times10^{-4})^2/1.59\times10^{-5} = 0.629\times10^{-3}$ ly/y$^2$, so we must have $X>C^2(r/v^2) = 1.59\times10^3$ ly. We will select a value about 5 times greater, which amazingly, is still within our own Galaxy.

As referenced in § 3.5, the author's 2-D simple Lorentzian (2-D SL) treatment was confirmed with Møller's GR-treatment for circular motion with a stationary observer A in the center and was also confirmed with Einstein's stationary observer E on the edge of a circle in Fig. 8. We will also use the same computer program that confirmed Einstein's original circular thought experiment with input of $v/C=1\times10^{-4}$, $r_E=1.59\times10^{-5}$ ly with DX=DY = -6.5$\times10^3$ ly or R=$(X^2+Y^2)^{1/2}$=9.2$\times10^3$ ly. The program normally starts a counter-clockwise, one-turn orbit at $\theta_0=0$ on the X-axis, but to begin the orbit at the beginning of the *approach-then-recede* mode, we begin the orbit at $\theta_0 = -0.136\ \pi$ which ends at $\theta_E=1.864\ \pi$ or $f=\theta/2\pi$ has range -0.068 to +0.932 as shown in Fig. 10. Both the predicted $\tau_F$ and $d\tau_F/dt$ from relativity theory are shown and the impossible region of $d\tau_F/dt <0$ extends from $f=-0.068$ to $f=+0.322$ which is somewhat more than the first quadrant as expected. Note also that the magnitude of maximum positive $d\tau_F/dt$ is significantly larger than the negative $d\tau_F/dt$ minimum just as in Fig. 6 compared to Fig. 7.

The swing in value of the predicted proper stationary time from +$\tau_F$ to -$\tau_F$ over one orbit as shown in Fig. 9 is proportional to distance X and becomes ridiculously large for distant galaxies. Even worse for time reversal, the rotational direction of the nearby stars must also oscillate over one orbit. So while astronomers and cosmologists have been lauding the merits of relativity theory and its cosmological constant, they have everyday, unknowingly, been providing data contrary to its predictions.

**4 CONCLUSIONS AND RECOMMENDATIONS**

*Time* is not equivalent to a spatial dimension and we have just seen that the incorporation of *time* with our three-dimensional space into the four-dimensional geometry of relativity theory has failed. However, as a predictive approximation, relative theory has been very fruitful. But now it is time to re-examine our basic physical concepts and search for a deeper understanding of our universe.

It is the Lorentz transformation that produced the unacceptable predictions above. Einstein (for SR) derived the Lorentz transformation from his two assumptions of the covariance of the laws of physics and the constancy of the velocity of light between two inertial frames and these concepts were carried as a local limit into his GR.

For the discussion and analyses to follow, let us characterize Einstein's noble effort as his pre-expansion attempt to force our universe locally into the Minkowski 4-D spacetime geometry and globally into the Riemann 4-D spacetime geometry. In view of his otherwise spectacular success, any attempt to remodel physical theory will certainly be a daunting (and initially unpopular) task.

It is seldom admitted but present physics still has no definitions or deeper understanding of our fundamental concepts of *space, time* or *energy* and no clear understanding of the source of *inertia* and *mass* or even why mass curves our 3-D space to produce *gravity*. To this list must be added the lack of understanding of why we must use quantum probabilities for the interactions of radiation and matter and why neither Einstein nor any theoretical physicist since has been able to unify relativity theory and quantum mechanics. Some fundamental dynamic is missing! A few comments follow on some of these physical infirmities.

**4.1 Spacetime**

The first conclusion is that both of Einstein's assumptions, (a) and (b), fail at sufficiently large distances even though they are excellent approximations locally and for everyday life. The Lorentz term that drives time backwards is the $vX/C^2$ coupling term of time dependence on spatial distance. Our concepts of space and time must be re-examined and we should search for mathematical definitions of these important concepts

**4.2 Electromagnetic Phenomena**

With the failure of relativity theory we have now returned to the very problem where Einstein began, that is, electromagnetic phenomena obey the Lorentz transformation but the mechanics of motion through space do not. We now have much more information about our universe than Einstein had, so perhaps nature is trying to tell us something very subtle about space and its expansion. On finer detail never before considered, possibly below the Planck scale, perhaps electromagnetics is also revealing a coexisting *medium* in our 3-D space that does indeed respond to its own different symmetric time. An additional fourth spatial dimensional support for our 3-D universe could open many possibilities.

**4.3 Space**

The Lorentz transformation may very well be compatible locally with electromagnetic phenomena and especially radiation, but on the global scale *space does not* fit the non-substantive-stretching-without-limit version of Einstein or present day physics. Real clocks do slow down locally as predicted by relativity theory but that prediction can also be obtained without that coupling term by just motion through a *substantive* space (*space* itself becomes a complex *ether*). The velocity of light would be the constant C locally at rest in a substantive space but at large distance r from the same origin in such an expanding space, the compounded radial velocity of light would be $v_c = Hr \pm C$.

**4.4 Time**

*Time* is <u>not</u> just Einstein's "reading of a nearby clock." *Time* is an abstract concept and there is no one *real time* that can be deduced from nature. On the scale of the duration of our lives, we are satisfied that time is asymmetric such that events that happen in the order a,b,c never happen again as c,b,a. However in present physics, time is just a symmetric parameter that is introduced into differential equations and can run either forward or backward. But when we finally discover what is driving the expansion of our universe, we should be able to define an asymmetric *cosmic time* that meets our subjective perceptions as well as our physics.

**4.5 Fundamental Concepts of Physics**

So this theoretical breakdown of relativity theory is a golden opportunity for physicists to pause and reflect on their fundamental conceptual building blocks that underlie all of physics.

      Some fundamental action, dynamic or dimension must be missing in our understanding of the machinery of nature which hopefully would eliminate the present outstanding problems of singularities, vacuum energy magnitude and composition of so-called "dark matter."

**4.6 Contributions of the Author**

The author has searched for the missing dynamic and sought the construction of a new model of our universe embracing such a new dynamic. This paper is not the proper venue to present that new dynamic or the resulting new cosmological model. However, references to some self-published details are (Leffert 1995; 1999; 2001a; 2001b; 2001c; 2001d). To summarize their contents: the model for the expansion of our universe is now essentially complete with no adjustable parameters. Its new definition of cosmic time allows the predicted cosmological parameters to agree with the astronomical evidence. Its new definition of space eliminates the infinite density singularities predicted by the present big bang theory. The model predicts the amount of and accounts for the nature of the dark mass and the source of the curvature of space and thus the source of gravity. Its new definition of energy accounts for the enormous vacuum energy predicted by particle physics and the very much smaller total energy density of the universe current in contemporary astronomy.

**5 ACKNOWLEDGEMENTS**
The author thanks his good friend, Emeritus Professor Robert A. Piccirelli, for extensive discussions of the new physical concepts.

**Table 1**
**Lorentz Transformation**
(Including constants δ… for translation of origins)

$$X = \gamma(x + vt + \delta x_o) \quad (1)$$
$$T = \gamma(t + vx/C^2 + \delta t_o) \quad (2)$$
$$x = \gamma(X - vT + \delta X_o) \quad (3)$$
$$t = \gamma(T - vX/C^2 + \delta T_o) \quad (4)$$
$$\gamma = 1/(1 - (v/C)^2)^{1/2} \quad (5)$$

**Table 2**
**Constants for Thought Experiments of Figures 1 and 3**
For A-B system: $DX = 0$
For F-B system: $DX = -2v_{B1}T_1$
$C=1$ ly/y, $T_2=12$ y, $T_1=T_2/2$, $v_{B2}=-v_{B1}$

<u>World Line B.1</u>
$V_{B1}/C=0.5$, $\gamma \approx 1.15$, $X_1 = v_{B1}T_1 = 3.0$ ly
$\delta x_o = +DX/\gamma$, $\delta t_0 = 0$
$\delta X_o = -DX$, $\delta T_o = +(v_{B1}/C^2)DX$

<u>World Line B.2</u>
$V_{B2}/C = -0.5$, $\gamma \approx 1.15$
$\delta x_o = (2X_1 + DX)/\gamma$ ($\approx 5.2$, $DX=0$), $\delta t_o = 0$
$\delta X_o = -2X_1 - DX$, $\delta T_o = -2(v_{B1}/C)^2 T_1 - (v_{B1}/C^2)DX$
For $DX=0$: $\delta X_o = -6.0$ ly, $\delta T_o = -3.0$ y

## Table 3
## Simple Lorentzian Treatment
Derivation from Lorentz Transformation of the Moving-Frame
Predicted Stationary Proper Clock A Reading $\tau_A$ and Position $x_A$ from Just the
Stationary Coordinates $(X_e, T_e)$ and Motion $(u, \dot{u})$ of the Moving Clock B

**Instantaneous x-axes (Ixa) or Lines of Simultaneity, t=constant**
    Let "e" represent any event on moving B's world line with
    B-coordinates $(x_e=0, t_e)$.
**Outbound**; $u=+|v|$ Let Ixa intercept stationary world line A
    At $(X_m=0, T_m=\tau_A)$ and B-coordinates are $(x_m, t_m)$
    From (4), $t_m=\gamma(T_m-|v|X_m/C^2) = \gamma T_m = \gamma \tau_A$ (constant)
    From (4), $\gamma \tau_A = \gamma(T_e-|v|X_e/C^2)$
    Or, $(\tau_A - T_e) = -|v|X_e/C^2$                                    (6)
    From (3), $x_m=\gamma(0-|v|T_m) = -\gamma|v|\tau_A$
    Using (6), $x_m=-\gamma|v|(T_e-|v|X_e/C^2)$
    From (3), for $x_e=0$, $T_e=X_e/|v|$
    So, $x_m=-\gamma X_e(1-v^2/C^2)=-X_e/\gamma$                   (7)
    Angle $\alpha$ from horizontal, $\alpha=\tan^{-1}[(T_e-\tau_A)/(X_e/C)=\tan^{-1}(+|v|/C)$
**Inbound**; $u=-|v|$ Let Ixa intercept stationary world line A
    At $X_n=0$, $T_n=\tau_A$ and B-coordinates are $(x_n, t_n)$
    From (4), $t_n=\gamma(T_n+|v|X_n/C^2-2(v/C)^2 T_e)=$
               $\gamma(T_n-2(v/C)^2 T_1=$constant$=\gamma(\tau_A-2(v/C)^2 T_1)$
    From (4), $\gamma(\tau_A-2(v/C)^2 T_1)= \gamma(T_e+|v|X_e/C^2-2(v/C)^2 T_1)$
    Or, $(\tau_A - T_e) = +|v|X_e/C^2$                                  (8)
    From (3), $x_n=\gamma(0+|v|T_n-2X_1)=+\gamma(|v|\tau_A-2X_1)$
    Using (8), $x_n = +\gamma|v|(T_e+|v|X_e/C^2-2X_1$
    From (3) for $x_e=0$ and $\delta X_o=-2X_1$, $T_e=-X_e/|v|+2X_1$
    So $x_n=-X_e/\gamma$ in agreement with (7) as it should be.       (9)
    Angle $\alpha$ from horizontal, $\alpha=\tan^{-1}[(T_e-\tau_A)/(X_e/C)=\tan^{-1}(-|v|/C)$
**In General;** For an arbitrary event "e" on B's world line where $u=\pm v$,
  The instantaneous x-axis at angle $\alpha$, intercepts A's stationary world line
    in terms of A-coordinates $(X_e, T_e, u)$ at:
        $\alpha = \tan^{-1}(u/C)$                                             (10)
        $\tau_A = T_e - (u/C)(X_e/C)$                           (11)
        $x_A = -X_e/\gamma$                                         (12)
**Derivative of (11) to Include Acceleration:**
    $d\tau_A/dt = \gamma d\tau_A/dT = \gamma(1 - (1/C^2)[(u\dot{X})+(X\dot{u})])$     (13)
In limit of uniform motion $\dot{u}=0$, $\dot{X}=u$ and $d\tau_A/dt = 1/\gamma>0$ for inertial systems.
**Hyperbolic Motion for Constant Acceleration:** (Constant force F):
    $g=F/m_o=d(\gamma \cdot u)/dt$         $u=dX/dt$
    $u = \sigma/(1+(\sigma/C)^2)^{1/2}$        $\sigma = gT + u_o\gamma_{uo} = \gamma u$     (14)
    $X=X_o+(C^2/g)\{[1+(1/C^2)(gT+\gamma_{uo}u_o)^2]^{1/2} - [1 + \gamma_{uo}^2(u_o/C)^2]^{1/2}\}$(15)

**Table 4**
**Two Dimensional Lorentzian Treatment for**
**Relativistic Circle Thought Experiment**
(Continuous Motion at Constant v where Angle θ is
Measured Counter Clockwise from X-axis)

$$X = R \cos(\theta) + X_o \tag{16}$$
$$Y = R \sin(\theta) + Y_o \tag{17}$$
$$u_X = -v \sin(\theta) \tag{18}$$
$$u_Y = +v \cos(\theta) \tag{19}$$
$$\tau_{Ax} = T - (u_x/C)(X/C) \tag{20}$$
$$\tau_{Ay} = T - (u_y/C)(Y/C) \tag{21}$$
$$\tau_A = (\tau_{Ax} + \tau_{Ay})/2 \tag{22}$$
$$\emptyset = \tan^{-1}(Y/X) - \theta + \pi/2 \tag{23}$$
$$w = \mathbf{u} \cdot \mathbf{R_p}/R_p = u\cos(\emptyset) \tag{24}$$
$$\gamma_r = 1/(1 - (w/C)^2)^{1/2} \tag{25}$$
$$r_A = -R_p/\gamma_r \tag{26}$$

Minimum translation $R_o = (X_o^2 + Y_o^2)^{1/2}$ of "at-rest" origin, $d\tau_A/dt = 0$ before
$d\tau_A/dt < 0$. For $|X_o| = |Y_o|$, $R_o = 2^{1/2}|X_o|$ and $X_o = Y_o = -2^{1/2}R/(v/C)^2$ (28)

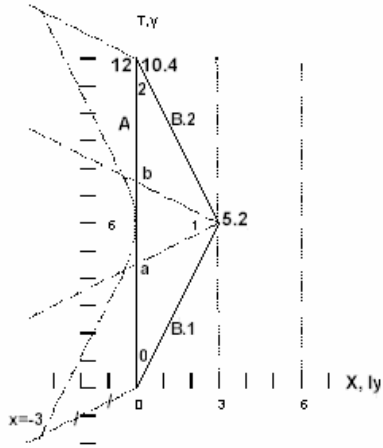

Fig. 1. A Minkowski diagram is shown for the standard clock paradox (CP) thought experiment at $|v/C|=0.5$ with assumed infinite acceleration at event "1". Observer B's retro-thruster is fired to reverse velocity at $\tau_B = \mathrm{tau}_B = T_1/\gamma = 5.20$ y. The lines joining events "1" to "a" and "1" to "b" represent the instantaneous x-axes for observer B at $\tau_B = 5.2$ y on the outbound leg B.1 and inbound leg B.2, respectively. Coordinate clock world lines are illustrated for A at X=3 ly and 6 ly and one for B at x = -3 ly.

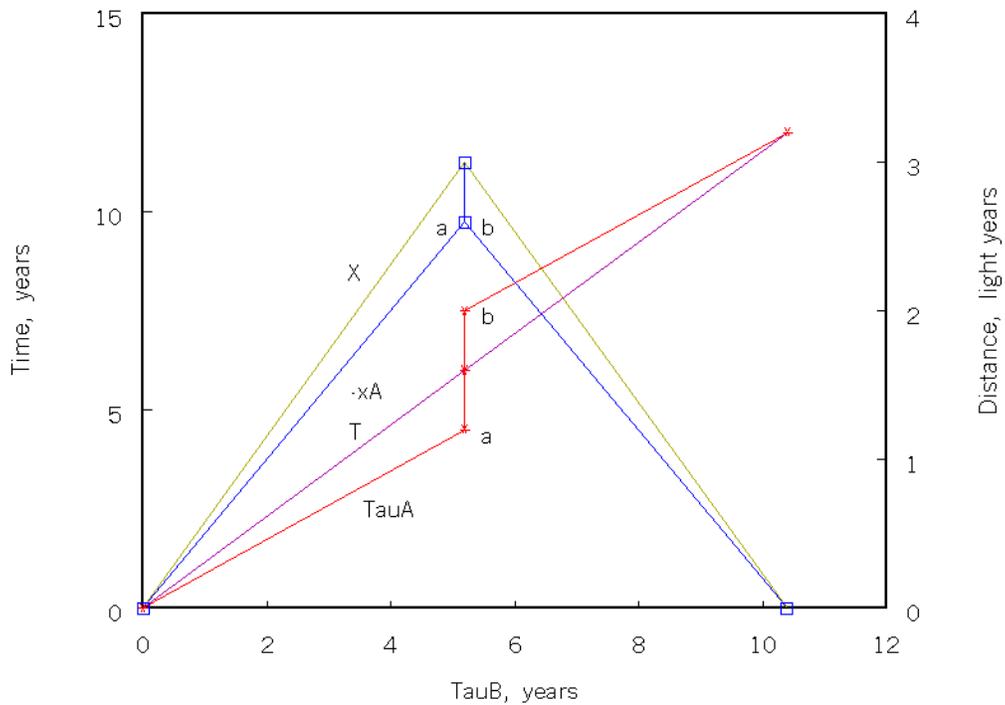

Fig. 2. For the solution of the standard CP thought experiment, predicted measurements by observers A (X,T) and B ($x_A$, $\tau_A$) are plotted versus the proper clock reading $\tau_B$ of observer B. The curves labeled $\tau_A$ and $x_A$ represent B's measurements of the reading and position of proper clock A. Note that $\tau_A = T$ and $x_A = -X$ only when A and B are together or at relative rest.

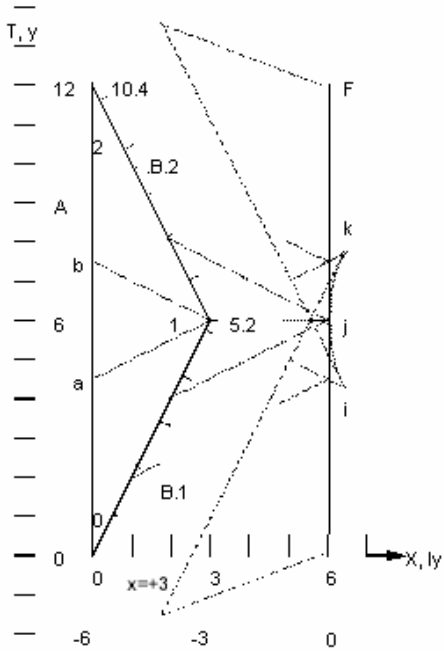

Fig. 3. For the modified CP thought experiment, observer A's friend F is introduced at rest relative to A at X=6.0 ly. Observer B's instantaneous x-axis at event "1" is extended to generate three new events "i", "$j_n$" and "k" on F's world line. Observer B continues measurements from the old origin, but F measures from the new F-origin. The unacceptable coordinate clock world line for x=+ 3 ly is dashed (See also Fig. 4).

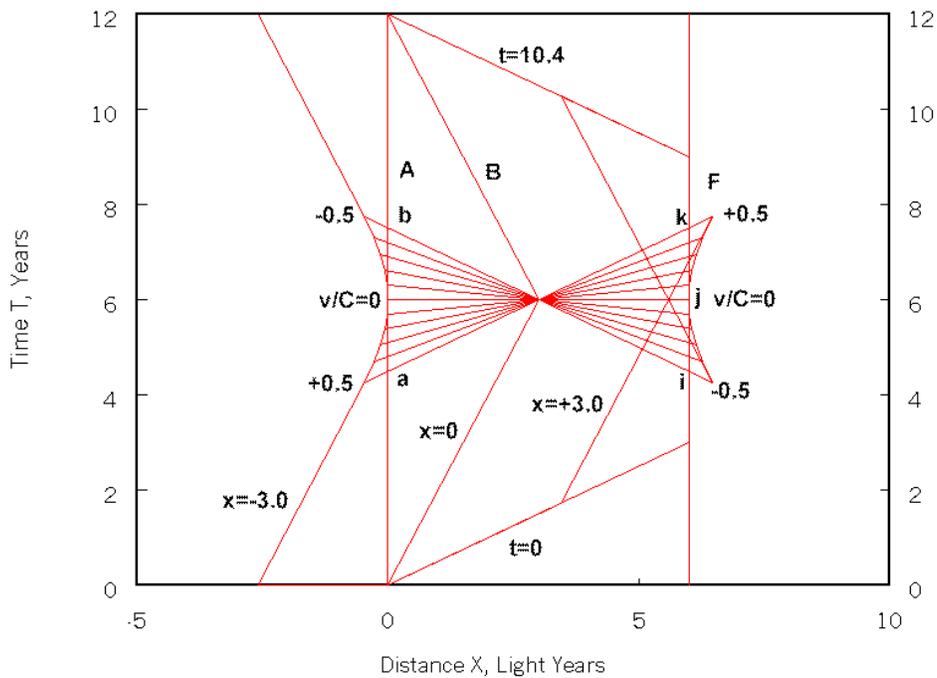

Fig. 4 A computer produced, Lorentz transformation, Minkowski diagram combining Figs. 1 and 3 to emphasize the clockwise rotation of moving observer B's instantaneous x-axis during B's turnaround. The impossible predicted world line of B's coordinate clock at x= +3 ly exemplifies how other such B-coordinate clock world lines conspire to predict F's proper clock to run backward.

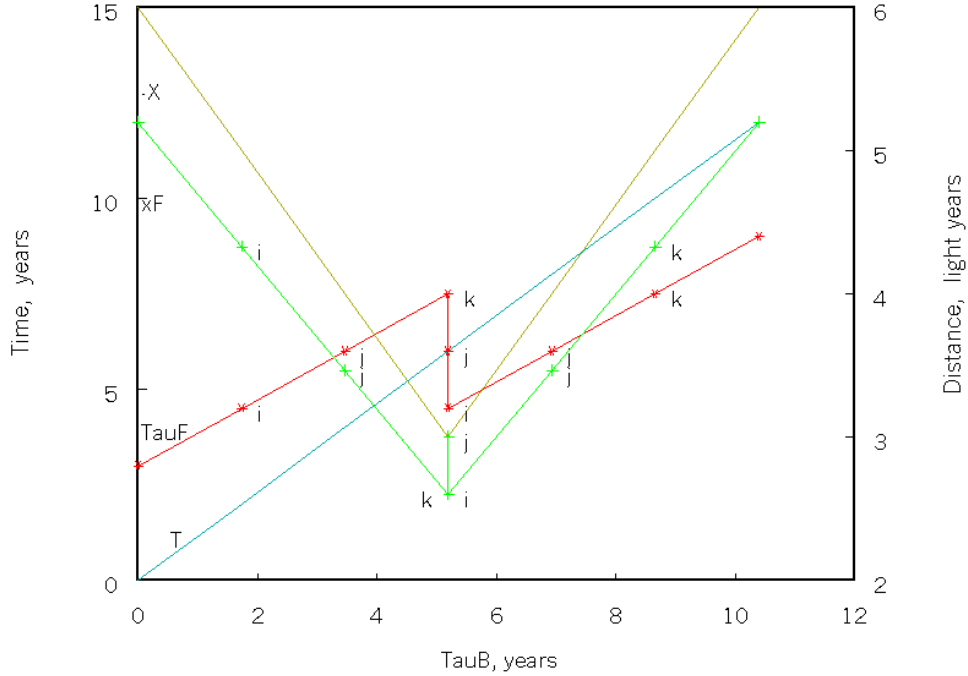

Fig. 5. The variables measured for the modified thought experiment are the same as in Fig. 2 but the predicted behavior is ominously different. Indeed, all events on the F world line between "i" and "k" occur at three different times for B and relativistic time for F is predicted to run backwards in the vertical section of $\tau_F$ which reverses the order of events "i", "$j_n$","k" and thus violates causality. Multiple values for B at event "1" are due to the limit of infinite turnaround acceleration, but the time reversal is inherent in the Lorentz transformation as will be shown.

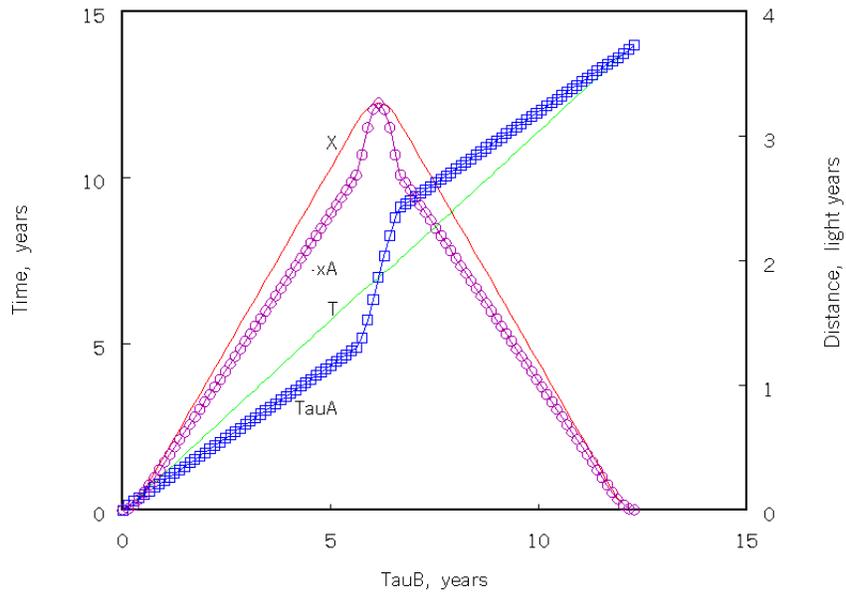

Fig. 6. For the standard clock paradox thought experiment, the SL-solution is shown with finite periods of acceleration. The variables measured are the same as those for Fig. 2. Besides the same $\Delta T=6$ y for each of the intervals of uniform motion at $|v/C|=0.5$, the four intervals of acceleration were of duration $\Delta T=0.5$ y which required accelerations $|g|=1.15$ ly/y2. The time duration of the entire trip for A was increased from $\Delta T=12$ to 14 y.

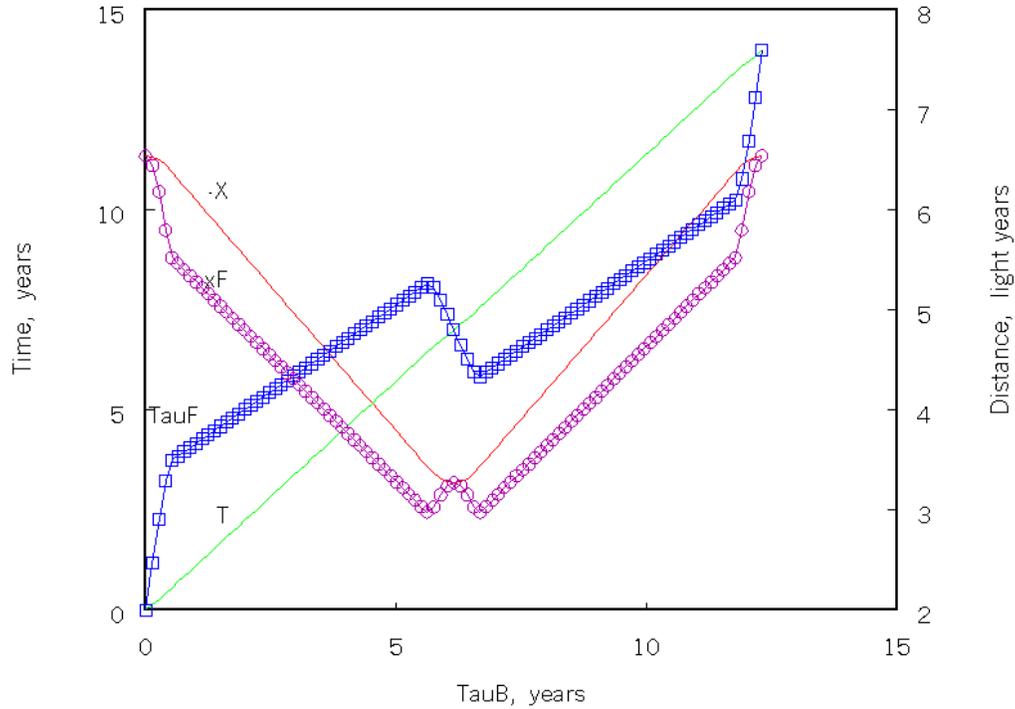

Fig. 7. For the modified thought experiment, the SL-treatment solution is shown with finite periods of acceleration. The variables measured are the same as those for Fig. 5. The computer program and input were exactly the same as for Fig. 6 except for one input for origin translation with $\Delta X = -6.0$ ly. Clearly the same unacceptable prediction of a measured time reversal of $\tau_F$ and violation of causality is demonstrated.

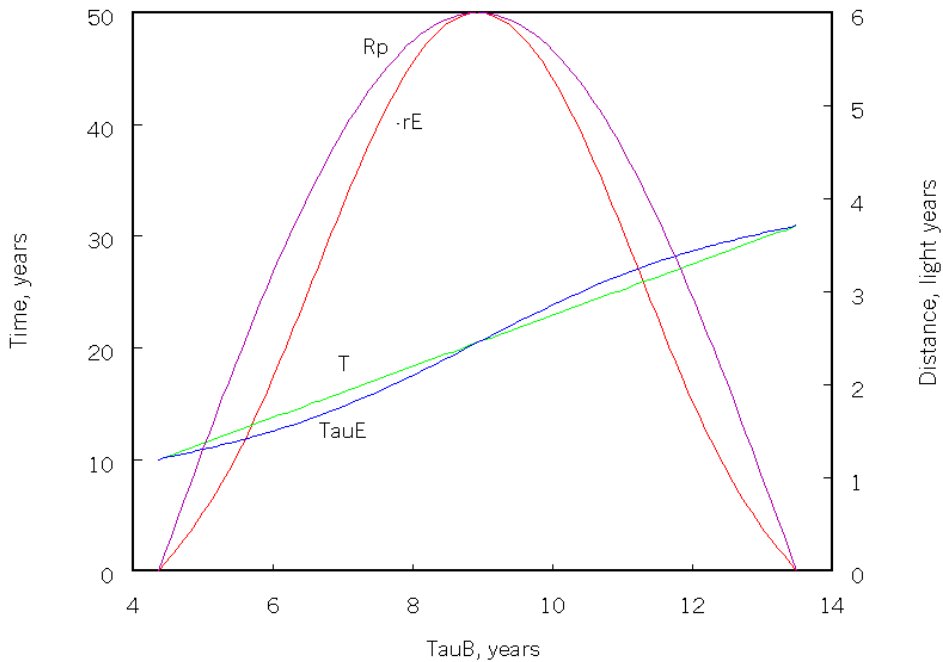

Fig. 8. The 2-D SL-treatment of Einstein's 1905 circular thought experiment has his friend E stationed on the edge of the circle at $X=R=3$ ly and the moving clock B is rotating at $v/C=0.9$ counter-clockwise during one rotation. His prediction that the B-clock period $\Delta t$ would lag E-clock $\Delta T$ by $[1-(1-(v/C)^2)^{1/2}]\Delta T$ is confirmed. Note radius measurements $R_p = -r_E$ at zero relative velocity at beginning, end and midpoint.

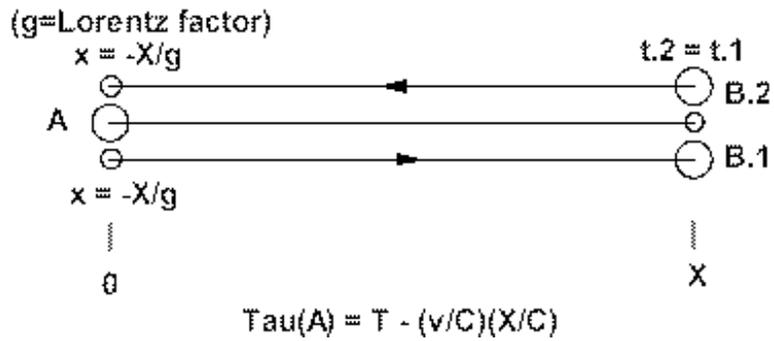

Fig. 9 . Here, in this revised version of the standard clock paradox of Fig. 1, a surrogate B-2 makes the return trip on a third framework of rods and clocks and is shown just as B-2 passes B-1 in the opposite direction. It is impossible for the photographs of the proper A clock by the B-1 and B-2 coordinate clocks to show the different readings of 4.5 y and 7.5 y as predicted by relativity theory in Fig. 1

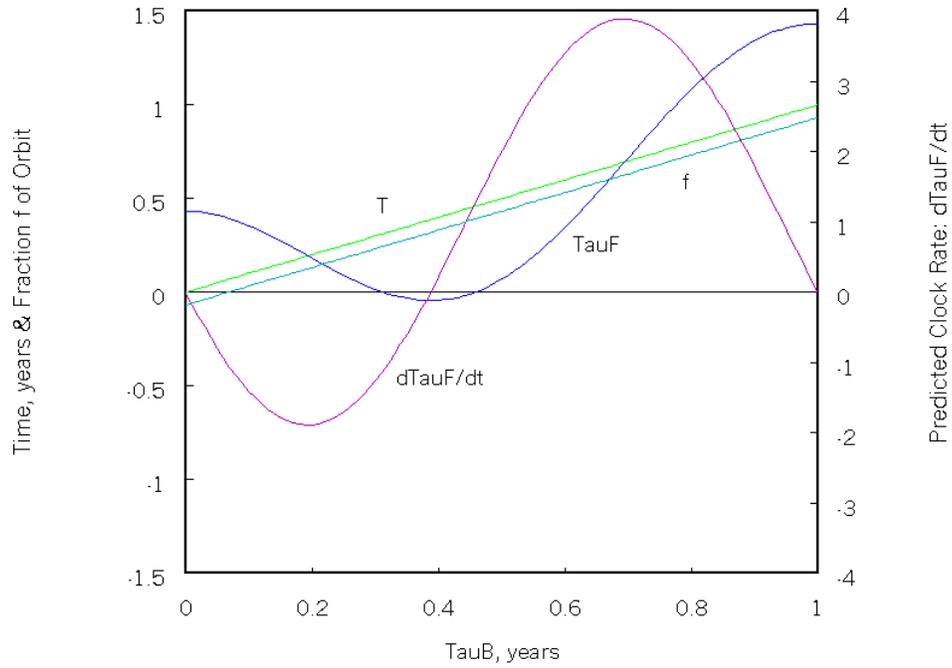

Fig. 10. During the beginning of this one orbit around the Sun, after adjustment for light travel time, relativity theory predicts astronomers would measure the impossible for a distant, 9.2 ly, active supernova Ia: that the supernova is evolving backwards, $d\tau_F/dt<0$, toward its original star. Not until f~0.322 does the direction of the nova time $\tau_F$ reverse and then it continues with normal evolution, $d\tau_F/dt>0$ until next cycle.